\documentclass[12pt]{amsart}
\usepackage{amsmath,amssymb,esint,enumerate,graphicx,hyperref}
\usepackage[sort&compress,numbers]{natbib}

\hypersetup{pdftex,colorlinks=true,allcolors=blue}
\usepackage[all]{hypcap}

\usepackage[foot]{amsaddr}
\usepackage[sc]{mathpazo}

\newcommand\ack{\subsection*{Acknowledgment}}

\DeclareMathAlphabet\mathsfbi{T1}{phv}{b}{it}

\numberwithin{equation}{section}

\newcommand\BV{\boldsymbol} 

\newcommand\dif{\:\!\mathrm{d}}

\newcommand\parderiv[2]{\frac{\partial #1}{\partial #2}}

\newcommand\Dderiv[2]{\frac{\mathrm D #1}{\mathrm D #2}}
\newcommand\Mach{\mathit{Ma}}
\newcommand\Rey{\mathit{Re}}
\newcommand\Pran{\mathit{Pr}}

\def\HS{{HS}}

\begin{document}

\author[Rafail V. Abramov]{Rafail V. Abramov}

\address{Department of Mathematics, Statistics and Computer Science,
University of Illinois at Chicago, 851 S. Morgan st., Chicago, IL 60607}

\email{abramov@uic.edu}

\title[A weakly compressible model of gas flow at low Mach number]
{Turbulence via intermolecular potential: A weakly compressible model
of gas flow at low Mach number}

\begin{abstract}
In our recent works we proposed a theory of turbulence in inertial gas
flow via the mean field effect of an intermolecular potential. We
found that, in inertial flow, turbulence indeed spontaneously develops
from a laminar initial condition, just as observed in nature and
experiments. However, we also found that density and temperature in
our inertial flow model behave unrealistically. The goal of the
current work is to demonstrate technical possibility of modeling
compressible, turbulent flow at low Mach number where both density and
temperature behave in a more realistic fashion. Here we focus on a new
treatment of the pressure variable, which constitutes a compromise
between compressible, incompressible and inertial flow. Similarly to
incompressible flow, the proposed equation for the pressure variable
is artificial, rather than derived directly from kinetic formulation.
However, unlike that for incompressible flow, our pressure equation
only damps the divergence of velocity, instead of setting it directly
to zero. We find that turbulence develops in our weakly compressible
model much like it does in the inertial flow model, but density and
temperature behave more realistically.
\end{abstract}

\maketitle

\section{Introduction}

Observations of turbulence in flows of fluids go at least as far back
as Leonardo da Vinci, however, the first mathematical (albeit somewhat
primitive) model of turbulent momentum transfer was proposed by
\citet{Bou}.  Six years later, \citet{Rey83} discovered that an
initially laminar flow of water in a straight smooth pipe
spontaneously develops turbulent motions whenever the high Reynolds
number condition is satisfied, even if all reasonable care is taken to
not disturb the flow artificially. Almost sixty years later,
Kolmogorov \citep{Kol41a,Kol41b,Kol41c} and \citet{Obu41} observed
that the time-averaged Fourier spectra of the kinetic energy of an
atmospheric wind possess universal decay structure, corresponding to
the inverse five-thirds power of the Fourier wavenumber. Numerous
attempts to explain the physical nature of turbulence have been made
throughout the twentieth century
\citep{Ric26,Tay,Tay38,KarHow38,Pra38,Obu49,Cha49,Cor51,Kol62,Obu62,Kra66a,Kra66b,Saf67,Saf70,Man74},
yet none successfully. So far, the process of spontaneous turbulence
formation in an initially laminar flow, as well as power scaling of
turbulent kinetic energy spectra, remained irreproducible, which
points to a deficiency in the conventional equations of fluid
mechanics. For an example, one can refer to relatively recent works by
\citet{AviMoxLozAviBarHof,BarSonMukLemAviHof} and
\citet{KhaAnwHasSan}, where turbulent-like motions in a numerically
simulated flow had to be created via deliberate perturbations.

In our recent works \citep{Abr22,Abr23,Abr24,Abr25} we proposed and
investigated a theory of turbulence in a gas, where turbulent motions
in an initially laminar, inertial (that is, constant pressure) flow
were created via the average effect of gas molecules interacting by
means of their intermolecular potential $\phi(r)$. In our theory, this
average effect is expressed via the corresponding mean field potential
$\bar\phi$, and constitutes a density-dependent correction to the
pressure gradient in the momentum equation.  This effect is absent
from the conventional Euler and Navier--Stokes equations of fluid
mechanics because, in the Boltzmann--Grad limit \citep{Gra}, it is
tacitly assumed that the effect of $\bar\phi$ is negligible. However,
we found that, in our model of inertial gas flow, turbulent motions
spontaneously emerge in the presence of $\bar\phi$, while failing to
do so in its absence~\cite{Abr24}.

In the past, a similar idea was proposed by \citet{Tsu}, who theorized
that turbulence was created via long-range correlations between
molecules, however, Tsug\'e's result was restricted to incompressible
flow. Yet, from what we discovered thus far, it appears that
turbulence spontaneously emerges from density fluctuations in an
initially laminar flow, which means that a model of turbulence must be
compressible. In our past work \citep{Abr20}, we also considered
long-range interactions as a possible reason for the manifestation of
turbulence.  However, we later observed that even the short-range hard
sphere potential creates turbulence in our model, which means that a
typical intermolecular potential (e.g. the Lennard-Jones potential
\citep{Len}) is also capable of the same effect.

At this stage, it is evident from our recent works
\citep{Abr22,Abr23,Abr24,Abr25}, that spontaneous turbulent motions in
an otherwise laminar flow are created by the mean field effect of the
intermolecular potential, which thus far was overlooked in the
conventional equations of fluid mechanics. However, the new,
subsequent, question has arisen in our study, which is how to treat
the pressure variable?  Conventional equations of fluid mechanics
offer two options: either compressible flow, where pressure has its
own transport equation, or incompressible flow, where it is
artificially set to keep the divergence of velocity at zero.
Unfortunately, neither of the two options seems to be compatible with
the mechanics of turbulence creation; as we show below, compressible
equations generally behave poorly at low Mach number (which is a well
known problem), while the structure of incompressible equations
precludes turbulence from developing. In our works
\citep{Abr22,Abr23,Abr24,Abr25} we suggested a third option of setting
the pressure to a constant (inertial flow). However, while turbulent
motions indeed develop spontaneously in our inertial flow, we found
that the dynamics of density and temperature can be unrealistic.

In the current work, we propose a novel treatment of the pressure
variable, which, by its design, constitutes a compromise between
compressible, incompressible and inertial flow. Similarly to
incompressible flow, the proposed pressure equation is artificial,
rather than derived directly from the kinetic formulation. However,
unlike that for incompressible flow, our pressure equation only damps
the divergence of velocity, instead of setting it directly to zero. In
the resulting model, turbulence spontaneously develops from a laminar
flow, and, at the same time, density and temperature behave more
realistically.

The work is organized as follows. In Section~\ref{sec:theory} we
propose a model for weakly compressible flow. In
Section~\ref{sec:comp_implementation} we describe the computational
implementation of the proposed model. In Section~\ref{sec:results} we
present the results of our numerical experiment with the proposed
model. Section~\ref{sec:summary} summarizes the results of the work.

\section{A new model of weakly compressible flow}
\label{sec:theory}

For convenience, here we start with introducing the advective (or
``material'') derivative operator for the flow velocity $\BV u$, given
via
\begin{equation}
\Dderiv ft\equiv\parderiv ft+\BV u\cdot\nabla f,
\end{equation}
whose geometric meaning is the directional derivative with respect to
time $t$ along the direction of the flow. With the help of this
operator, the transport equations for the density $\rho$ and velocity
$\BV u$ are written, respectively, via
\begin{equation}
\label{eq:rho_u_gen}
\Dderiv\rho t=-\rho\nabla\cdot\BV u,\qquad\rho\Dderiv{\BV u}t=-\nabla
(p+\bar\phi)+\nabla\cdot(\mu\nabla\BV u),
\end{equation}
where $p$, $\mu$ and $\bar\phi$ are the pressure, viscosity and the
mean field potential, respectively. The latter has been introduced in
our recent works \citep{Abr22,Abr23,Abr24,Abr25}, and is not a part of
the conventional fluid mechanics equations. The general formula for
the mean field potential $\bar\phi$ is given \citep{Abr24} via
\begin{equation}
\label{eq:bphi}
\bar\phi=\frac{2\pi p\rho}{3m}\int_0^\infty\big(1-e^{-\phi(r)/\theta
}\big)\parderiv{}r\big(r^3Y(r)\big)\dif r,
\end{equation}
where $m$ is the mass of a gas molecule, $\phi(r)$ is the short-range
intermolecular potential, $\theta=p/\rho$ is the kinetic temperature,
and $Y(r)$ is the cavity distribution function \citep{Bou06} for a
pair of molecules. Under the assumption that the gas is sufficiently
dilute (i.e.  $Y\approx 1$), and that the intermolecular potential
$\phi(r)$ can be approximated via the hard sphere potential $\phi_\HS$
with effective range $\sigma$, that is,
\begin{equation}
\phi_\HS(r)=\left\{\begin{array}{l@{,\qquad}l@{,}} \infty & r\leq
\sigma \\ 0 & r>\sigma\end{array}\right.
\end{equation}
the general formula in \eqref{eq:bphi} simplifies to
\begin{equation}
\label{eq:bphi_HS}
\bar\phi_\HS=\frac{4\rho p}{\rho_\HS}.
\end{equation}
Above, $\rho_\HS=6 m/\pi\sigma^3$ is the density of an equivalent hard
sphere of mass $m$ and diameter $\sigma$. Upon substitution of
\eqref{eq:bphi_HS} into \eqref{eq:rho_u_gen}, we arrive at
\begin{equation}
\label{eq:rho_u}
\Dderiv\rho t=-\rho\nabla\cdot\BV u,\qquad\rho\Dderiv{\BV u}t=-\nabla
\left[p\left(1+\frac{4\rho}{\rho_\HS}\right)\right]+\nabla\cdot(\mu
\nabla\BV u).
\end{equation}
Thus far, the pressure $p$ is the only variable which lacks an
equation. This is intentional; while the transport equations for the
density and velocity variables are universal, the treatment of the
pressure variable depends on the dynamical regime of the flow.

In compressible flow at high Mach number, the pressure $p$ is equipped
with its own transport equation, which is derived from the Boltzmann
equation by computing velocity moments of the distribution function
\citep{CerIllPul,HirCurBir}:
\begin{equation}
\label{eq:p_compressible}
\frac 1\gamma\Dderiv pt=-p\nabla\cdot\BV u+\nabla
\cdot\left(\frac\mu\Pran\nabla\theta\right).
\end{equation}
Above, $\gamma$ and $\Pran$ are the adiabatic exponent and the Prandtl
number, respectively, and we also forgo the stress term for
simplicity.  The equations in \eqref{eq:rho_u} (with $\rho_\HS$ set to
infinity) and \eqref{eq:p_compressible}, taken together as a system,
comprise the standard compressible Navier--Stokes equations
\citep{Bat,Gols}. However, it is well known that these compressible
equations do not constitute an accurate model of compressible flow at
low Mach and high Reynolds numbers. In particular, if one attempts to
model such a flow via a numerical simulation, the resulting numerical
solution will be inundated with unrealistic amounts of acoustic waves,
which clearly does not happen in nature.

\subsection{Rescaling and non-dimensionalization}

The reason for inaccurate behavior of the system in \eqref{eq:rho_u}
and \eqref{eq:p_compressible} at low Mach number can be seen with the
help of a standard rescaling of variables. To this end, we introduce
reference constants $L$ and $U$, which correspond to the spatial scale
of the problem, and the characteristic speed of the flow,
respectively. We then rescale the variables $\BV x$, $t$ and $\BV u$
as follows:
\begin{equation}
\BV x=L\BV{\tilde x},\qquad t=\frac LU\tilde t,\qquad\BV u=U\BV{\tilde
  u}.
\end{equation}
It is easy to see that the advective derivative is rescaled via
\begin{equation}
\Dderiv ft=\frac UL\Dderiv f{\tilde t}.
\end{equation}
In addition, we introduce reference density $\rho_0$, temperature
$T_0$ (in degrees K), and viscosity $\mu_0$. The corresponding
rescalings for $\rho$, $p$ and $\mu$ are given via
\begin{equation}
\rho=\rho_0\tilde\rho,\qquad p=\rho_0\theta_0\tilde p,\qquad
\mu=\mu_0\tilde\mu,\qquad\theta_0=\frac{RT_0}M,
\end{equation}
where $R$ and $M$ are the universal gas constant and the molar mass of
gas, respectively. Upon substitution into \eqref{eq:rho_u} and
\eqref{eq:p_compressible}, we obtain the following non-dimensional
equations:
\begin{subequations}
\label{eq:rho_u_p_rescaled}
\begin{equation}
\Dderiv{\tilde\rho}{\tilde t}=-\tilde\rho\tilde\nabla\cdot\BV{\tilde u
},\qquad\tilde\rho\Dderiv{\BV{\tilde u}}{\tilde t}=-\frac 1{\Mach^2}
\tilde\nabla\tilde p-\frac{4\eta}{\Mach^2}\tilde\nabla(\tilde\rho
\tilde p)+\frac 1\Rey\tilde\nabla\cdot(\tilde\mu\tilde\nabla
\BV{\tilde u}),
\end{equation}
\begin{equation}
\label{eq:p_rescaled}
\frac 1\gamma\Dderiv{\tilde p}{\tilde t}=-\tilde p\tilde\nabla\cdot
\BV{\tilde u}+\frac 1\Rey\tilde\nabla\cdot\left(\frac{\tilde\mu}\Pran
\tilde\nabla\tilde\theta\right).
\end{equation}
\end{subequations}
Above, $\Mach$, $\Rey$ and $\eta$ are, respectively, the Mach number,
the Reynolds number, and the packing fraction:
\begin{equation}
\Mach=\frac U{\sqrt{\theta_0}},\qquad\Rey=\frac{\rho_0UL}{\mu_0},
\qquad\eta=\frac{\rho_0}{\rho_\HS}.
\end{equation}
At normal conditions, the packing fraction $\eta\sim 10^{-3}$
\citep{Abr23}. As a result, for typical low Mach number values, say,
$\Mach\sim 0.1$ (which corresponds to $U\sim$ 30 m/s at normal
conditions), we have $\Mach^2\ll 1$, while $4\eta/\Mach^2\sim 1$.
Additionally, for a flow to be turbulent, $\Rey\gg 1$ (see
\citet{Rey83}).  It is thus apparent that, at low Mach numbers, the
term with the pressure gradient dominates the momentum equation, if
the non-dimensional pressure gradient itself is $\tilde\nabla\tilde
p\sim 1$.

However, it is known from observations that gas flows at normal
conditions and low Mach number do not exhibit any irregular
behavior. This, in turn, means that, in order for the rescaled
equations in \eqref{eq:rho_u_p_rescaled} to accurately model real gas
flows, the non-dimensional pressure gradient must at least satisfy
$\tilde\nabla\tilde p\lesssim O(\Mach^2)$. However, at high Reynolds
number, it can only happen, at least on the non-dimensional time scale
$\tilde t\sim 1$, if $\tilde\nabla\cdot \BV{\tilde u}\sim O(\Mach^2)$.
Irrespectively of physics, there are two principal technical
approaches to control the gradient of the pressure:
\begin{itemize}
\item {\bf The direct approach:} explicitly set the pressure gradient
  to a desired value (such as zero);
\item {\bf The indirect approach:} implicitly control the pressure
  gradient by appropriately restricting the magnitude of velocity
  divergence.
\end{itemize}
We already tested the direct approach in our recent works
\citep{Abr22,Abr23,Abr24,Abr25} by setting the pressure variable to a
constant, thus modeling inertial flow. Although this, rather
primitive, model indeed reproduces main features of turbulent flow,
there are some aspects of it which are unrealistic -- in particular,
variations of density and temperature are much larger than what is
typically observed.

A form of the indirect approach is used in the conventional equations
for incompressible flow, where the divergence of flow velocity is
preserved along a streamline via an artificial choice of the pressure
equation. Coupled with a divergence-free initial condition, the
equations for incompressible flow produce solutions which tend to
behave regularly at low Mach number. However, below we show that this
approach is unsuitable for our purpose, because turbulent dynamics are
not retained in the equations for incompressible flow.

\subsection{Incompressible flow}

A conventional way to control the divergence of velocity at low Mach
number is to replace the pressure transport equation in
\eqref{eq:p_compressible} with an artificial constitutive relation,
which deliberately sets the pressure in such a way so as to preserve
the divergence of velocity along a streamline, which leads to the
incompressible Navier--Stokes equations \citep{Bat}. The same
procedure can in fact also be used for our momentum equation in
\eqref{eq:rho_u} with the additional mean field potential term. In
order to proceed, first observe that, if we somehow ensure that
$\nabla\cdot\BV u=0$, then constant density $\rho=\rho_0$ becomes a
valid solution. Thus, presuming that the density variable is constant,
we can write the momentum equation in \eqref{eq:rho_u} in the form
\begin{equation}
\rho_0\Dderiv{\BV u}t=-(1+4\eta)\nabla p+\nabla\cdot(\mu\nabla\BV
u).
\end{equation}
Let us now compute the divergence of both sides of the equation above,
and, in addition, presume that viscosity $\mu=\mu_0$ is also constant:
\begin{equation}
\rho_0\nabla\cdot\Dderiv{\BV u}t=-(1+4\eta)\Delta p+\mu_0\Delta(
\nabla\cdot\BV u).
\end{equation}
Next, observe that
\begin{equation}
\nabla\cdot\Dderiv{\BV u}t=\parderiv{(\nabla\cdot\BV u)}t+\nabla\cdot(
\BV u\cdot\nabla\BV u)=\Dderiv{(\nabla\cdot\BV u)}t+\nabla\BV u^T:
\nabla\BV u,
\end{equation}
where ``$:$'' denotes the Frobenius product of two matrices. This
leads to
\begin{equation}
\rho_0\left(\Dderiv{(\nabla\cdot\BV u)}t+\nabla\BV u^T:\nabla\BV
u\right)=-(1+4\eta)\Delta p+\mu_0\Delta(\nabla\cdot\BV u).
\end{equation}
It is now clear that, in order to preserve $\nabla\cdot\BV u=0$ along
the streamline, the pressure must satisfy
\begin{equation}
\Delta p=-\frac{\rho_0}{1+4\eta}\nabla\BV u^T:\nabla\BV u.
\end{equation}
This is the artificial pressure equation in the incompressible
setting. In the conventional formulation without the mean field
potential forcing, the packing fraction $\eta$ is not present.
However, there is otherwise no difference in the dynamics -- clearly,
the velocity solution is the same for the same initial and boundary
conditions, while the pressure variable functions as a ``Lagrange
multiplier'' to ensure that the velocity field is divergence-free.
Thus, the presence of the mean field potential does not affect the
incompressible dynamics, and, therefore, does not produce turbulent
flow.

\subsection{New idea: weakly compressible flow}

In the current work, we explore a more flexible approach, where the
artificial pressure condition is chosen to constitute ``middle
ground'' between the incompressible and inertial regimes. We refer to
our new model as {\em weakly compressible flow}, because it aims to
control the divergence of velocity by damping it, rather than setting
it directly to zero.

Let us divide the momentum equation in \eqref{eq:rho_u} by $\rho$, and
compute the divergence on both sides:
\begin{equation}
\Dderiv{(\nabla\cdot\BV u)}t+\nabla\BV u^T:\nabla\BV u=-\nabla\cdot
\left(\frac 1\rho\nabla\left[p\left(1+\frac{4\rho}{\rho_\HS}\right)
  \right]\right)+\nabla\cdot\left(\rho^{-1}\nabla\cdot(\mu\nabla\BV
u)\right).
\end{equation}
After the rescaling and subsequent non-dimensionalization of the
variables, the equation above becomes
\begin{equation}
\label{eq:u_rescaled}
\Dderiv{(\tilde\nabla\cdot\BV{\tilde u})}{\tilde t}+\tilde\nabla \BV{
  \tilde u}^T:\tilde\nabla\BV{\tilde u}=-\frac 1{\Mach^2}\nabla\cdot
\left(\tilde\rho^{-1}\nabla\left[\tilde p(1+4\eta\tilde\rho)\right]
\right)+\frac 1\Rey\tilde\nabla\cdot\left(\tilde\rho^{-1}\tilde\nabla
\cdot(\tilde\mu\tilde\nabla\BV{\tilde u})\right).
\end{equation}
Here, we propose to control the divergence of velocity by imposing
linear damping on it via the following artificial pressure equation:
\begin{equation}
\label{eq:p_new_rescaled}
\tilde\nabla\cdot\left(\tilde\rho^{-1}\tilde\nabla\tilde p
\right)=\tilde\tau^{-1}\tilde\nabla\cdot\BV{\tilde u},
\end{equation}
where $\tilde\tau\sim 1$ is the non-dimensional characteristic time of
decay, chosen empirically. Then, the balance of terms in the
right-hand side of \eqref{eq:u_rescaled} is given via
\begin{equation}
\Dderiv{(\tilde\nabla\cdot\BV{\tilde u})}{\tilde t}+\tilde\nabla\BV{
  \tilde u}^T:\tilde\nabla\BV{\tilde u}=-\frac{\tilde\nabla\cdot\BV{
    \tilde u}}{\Mach^2\tilde\tau}+O\left(\frac{4\eta}{\Mach^2}\right)
+O\left(\Rey^{-1}\right),
\end{equation}
that is, such substitution renders $\tilde\nabla\cdot\BV{\tilde u}\sim
\Mach^2$, at least on the non-dimensional time scale $\tilde t\sim
1$. Generally, to control the divergence, the right-hand side of
\eqref{eq:p_new_rescaled} can be a linear (or even nonlinear) positive
definite operator acting on $\tilde\nabla \cdot\BV{\tilde u}$,
however, here we choose the simplest option in the form of linear
damping. As we can see, the proposed pressure equation in
\eqref{eq:p_new_rescaled} is completely artificial, just as the one
for incompressible flow. However, in our case, the flow remains
compressible (even if weakly), and thus still retains the turbulent
effect induced by the mean field potential.

\subsection{Simplified weakly compressible equations}

Reverting back to dimensional variables, and expressing the advection
terms in a conservative form, we arrive at
\begin{subequations}
\begin{equation}
\parderiv\rho t+\nabla\cdot(\rho\BV u)=0,\qquad\nabla \cdot\left(
\frac\tau\rho\nabla p\right)=\nabla\cdot\BV u,\qquad
\tau=\frac{UL}{\theta_0}\tilde\tau,
\end{equation}
\begin{equation}
\parderiv{(\rho\BV u)}t+\nabla\cdot(\rho\BV u^2)+\left(1+\frac{4\rho}{
  \rho_\HS}\right)\nabla p+\frac{4p}{\rho_\HS}\nabla\rho=\nabla\cdot
(\mu\nabla\BV u).
\end{equation}
\end{subequations}
At normal conditions, observe that $\rho/\rho_\HS\sim 10^{-3}$, which
is a small correction to unity. Thus, we discard this small correction
in the momentum equation, by setting
\begin{equation}
1+4\rho/\rho_\HS\to 1.
\end{equation}
Additionally, noting that, at low Mach numbers, the pressure consists
largely of its background state $p_0$ with small fluctuations, we
replace $p$ with $p_0$ in the density gradient term of the momentum
equation. The resulting simplified, weakly compressible equations are
\begin{subequations}
\label{eq:weakly_compressible}
\begin{equation}
\label{eq:rho_p}
\parderiv\rho t+\nabla\cdot(\rho\BV u)=0,\qquad\nabla \cdot\left(\frac
\tau\rho\nabla p\right)=\nabla\cdot\BV u,\qquad\tau=\frac{UL}{\theta_0
}\tilde\tau,
\end{equation}
\begin{equation}
\label{eq:u}
\parderiv{(\rho\BV u)}t+\nabla\cdot(\rho\BV u^2)+\nabla p+\frac{4 p_0
}{\rho_\HS}\nabla\rho=\nabla\cdot (\mu\nabla\BV u).
\end{equation}
\end{subequations}
Note that, as a result of the simplifications above, the nonlinear
coupling between $\rho$ and $p$ in the momentum equation has been
eliminated. Also, observe that one can obtain the inertial flow system
used in our recent works \citep{Abr22,Abr23,Abr24,Abr25} by setting
$\nabla p=\BV 0$ in \eqref{eq:u} and discarding the pressure equation
from \eqref{eq:rho_p}.

\section{Computational implementation of the model}
\label{sec:comp_implementation}

Here we describe the computational implementation of the simplified
weakly compressible equations in \eqref{eq:weakly_compressible}. As in
our recent works \citep{Abr22,Abr23,Abr24,Abr25}, we use
\textit{OpenFOAM}\footnote{\url{https://openfoam.org};
  \url{https://openfoam.com}} \citep{WelTabJasFur} to implement the
computational code.

\subsection{The time-stepping scheme}

The spatial discretization of \eqref{eq:weakly_compressible} is
implemented in \textit{OpenFOAM} via the standard second-order finite
volume scheme using the van Leer flux limiter \citep{vanLee}. However,
the time discretization is somewhat nontrivial, because $\nabla p$ in
\eqref{eq:u} effectively constitutes a linear damping of momentum. It
is well known that dissipative linear terms have to be treated
implicitly during time stepping to avoid numerical instability. Here,
we use a first order stepping with the time step denoted via $\Delta
t$,
\begin{subequations}
\label{eq:weakly_compressible_discretized}
\begin{equation}
\label{eq:rho_p_d}
\frac{\rho_*-\rho}{\Delta t}+\nabla\cdot(\rho\BV u)=0,\qquad\nabla
\cdot\left(\frac \tau{\rho_*}\nabla p_*\right)=\nabla\cdot\BV u_*,
\end{equation}
\begin{equation}
\label{eq:u_d}
\frac{\rho_*\BV u_*-\rho\BV u}{\Delta t}+\nabla\cdot(\rho\BV u\BV u_*)
+\nabla p_*+\frac{4 p_0 }{\rho_\HS}\nabla\rho_*=\nabla\cdot(\mu\nabla
\BV u_*),
\end{equation}
\end{subequations}
where the variables subscripted with asterisks refer to the next time
level. Above, observe that $\rho_*$ can be computed explicitly from
the density equation in \eqref{eq:rho_p_d}, while $p_*$ and $\BV u_*$
are coupled into a linear system of equations for pressure in
\eqref{eq:rho_p_d} and momentum in \eqref{eq:u_d}. We solve this
system of equations via an underrelaxed fixed point iteration for the
main diagonal of the discretized momentum equation (effectively
amounting to a Jacobi iteration). At each time step, the computation
proceeds as follows:
\begin{enumerate}[\indent (i)]
\item $\rho_*$ is evaluated from the density equation in
  \eqref{eq:rho_p_d};
\item $\BV u_*^{(1)}$ is evaluated from the momentum equation in
  \eqref{eq:u_d} with $p_*^{(0)}=p$;
\item $p_*^{(k)}$ is evaluated from the pressure equation in
  \eqref{eq:rho_p_d} for the current iterate $u_*^{(k)}$;
\item $\BV u_*^{(k+1)}$ is evaluated from an underrelaxed, diagonally
  dominant rearrangement of the momentum equation in \eqref{eq:u_d}
  for the current iterate $p_*^{(k)}$;
\item Steps (iii) and (iv) are repeated until the pressure equation in
  \eqref{eq:rho_p_d} becomes identity for the iterates $p_*^{(k)}$ and
  $\BV u_*^{(k)}$, subject to a suitable tolerance. Then, $p_*^{(k)}$
  and $\BV u_*^{(k)}$ are accepted as $p_*$ and $\BV u_*$,
  respectively, for the next time step.
\end{enumerate}
In the context of \textit{OpenFOAM}, the steps above are programmed
similarly to the implementation of the PISO algorithm~\citep{Iss} in
the standard \textit{icoFoam} solver.

\subsection{Description of the computational domain and parameters}

\begin{figure}
\includegraphics[width=\textwidth]{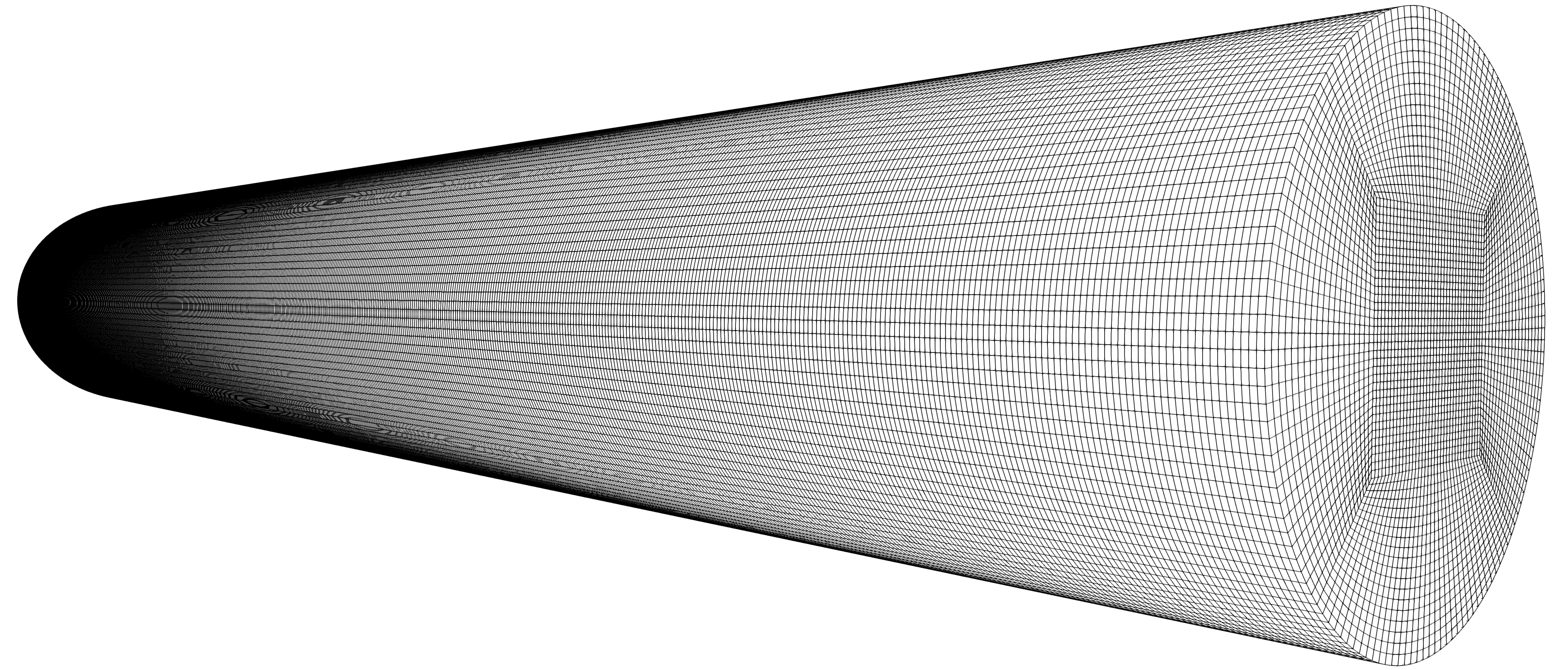}
\caption{General structure of the finite-volume discretization mesh of
  the cylindrical pipe.}
\label{fig:domain}
\end{figure}

In our recent works \citep{Abr22,Abr24,Abr25}, the computational
domain was a straight pipe of square cross-section, whose spatial
discretization consisted of uniform cubes. In the current work, we
``upgrade'' the domain to be a cylindrical pipe instead, which is
easily achievable via the finite-volume discretization, naturally
provided by \textit{OpenFOAM}. The pipe has 48~cm in length, and 7~cm
in diameter. The three-dimensional ``wire mesh'' of the pipe, which
illustrates the spatial discretization, is displayed in
Figure~\ref{fig:domain}.

\begin{figure}
\includegraphics[width=0.5\textwidth]{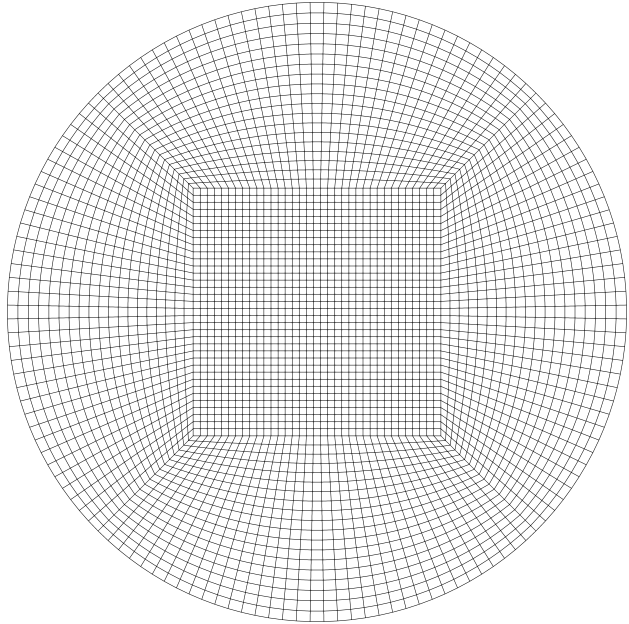}
\caption{The structure of the spatial discretization mesh in the
  transverse plane of the pipe.}
\label{fig:inlet}
\end{figure}

In the longitudinal direction, the spatial discretization is uniform
with step 0.8 mm, same as in our recent work \citep{Abr24}, which,
given the length of the pipe, yields 600 spatial discretization steps
along the axis of the pipe. The discretization scheme in the
transversal plane is shown separately in Figure~\ref{fig:inlet}; it
consists of the central 2.8$\times$2.8 cm$^2$ ``core'', with four
adjacent quadrangular blocks, whose outer circular edges form the
boundary of the pipe. The size of a discretization cell within the
central core is 0.8$\times$0.8 mm$^2$, such that the core has uniform
spatial discretization in all three dimensions. In the transversal
plane, the central core contains 35$\times$35=1225 cells, while each
of the four adjacent quadrangular blocks contains 35$\times$19=665
cells. Thus, the transversal plane contains 3885 cells in total, and
the total number of cells in the domain is 3885$\times$600$=$2 331
000.

\subsection{Initial and boundary conditions, and other parameters of
the simulation}

In the current work, we simulate air flow at normal conditions.  Thus,
the background pressure and the hard sphere density are set to
$p_0=101.3$ kPa and $\rho_\HS=1850$ kg/m$^3$ \citep{Abr23},
respectively.  The scaling parameters in \eqref{eq:rho_p} are set to
$U=$ 30 m/s (the maximum speed difference in the jet), $L=$ 1 cm (the
thickness of the jet) and $T_0=$ 293.15 K (20 $^\circ$C). The molar
mass of air and the universal gas constant are set to $M=$ 28.97 g/mol
and $R=$ 8.31446 kg m$^2$/s$^2$ mol K, respectively.

The resulting reference kinetic temperature $\theta_0$ and Mach number
$\Mach$ are given via
\begin{equation}
\theta_0=\frac{RT_0}M\approx 8.4\cdot 10^4\text{ m}^2/\text{s}^2,
\qquad\Mach=\frac U{\sqrt{\theta_0}}\approx 0.1,
\end{equation}
and we choose the non-dimensional empirical characteristic decay time
to be $\tilde\tau=$ 2.5. The viscosity variable $\mu$ is set to
\begin{equation}
\mu=\mu_0\sqrt{\frac p{\theta_0\rho}},
\end{equation}
where the reference value of viscosity is set to $\mu_0=1.825\cdot
10^{-5}$ kg/m s, to match the viscosity of air at 293.15 K.

The boundary conditions of flow in the pipe are chosen similarly to
those in our recent work \citep{Abr24}. Namely, the front end of the
pipe is treated as a wall, which houses a round axisymmetric inlet of
1 cm in diameter.  The back end of the pipe represents the outlet of
the flow.  The model variables are specified at the boundaries as
follows:
\begin{itemize}
\item {\bf Inlet and walls:} The Dirichlet boundary condition is set
  for velocity and temperature. The velocity variable has zero value
  at the walls, and a radially symmetric parabolic profile at the
  inlet, with the maximum value of 30 m/s at the center (as in the
  experiment of \citet{BucVel}), directed along the longitudinal axis
  of the pipe.  The temperature variable is set to 293.15 K at the
  inlet and the walls. The Neumann boundary condition (zero normal
  derivative) is specified for pressure.
\item {\bf Outlet:} For velocity and temperature, we set the Neumann
  boundary condition with zero normal derivative. The pressure
  variable is set to 101.3 kPa, which constitutes the Dirichlet
  boundary condition.
\end{itemize}
The initial state of the flow within the pipe is set as follows: the
temperature and pressure variables are prescribed constant values of
293.15 K and 101.3 kPa, respectively, while the velocity variable is
set to the laminar axisymmetric jet with parabolic profile matching
the inlet boundary condition, which extends throughout the length of
the pipe.

\section{Results of the numerical simulation}
\label{sec:results}

Here we report the results of the numerical simulation of air flow
within the cylindrical pipe, described above. What follows is
separated into two parts: first, we demonstrate how turbulent flow
develops from a laminar initial condition, and, second, we present the
properties of fully developed turbulent flow. The snapshots of various
quantities in
Figures~\ref{fig:velocity_t0.01-0.05}--\ref{fig:rho_T_p_t0.15} are
produced with help of
\textit{ParaView}\footnote{\url{https://paraview.org}}, while the
kinetic energy spectra in Figure \ref{fig:energy_spectra} are computed
and visualized using
\textit{Octave}\footnote{\url{https://octave.org}}.

\subsection{Development of turbulent flow from a laminar initial condition}

\begin{figure}
\includegraphics[width=\textwidth]{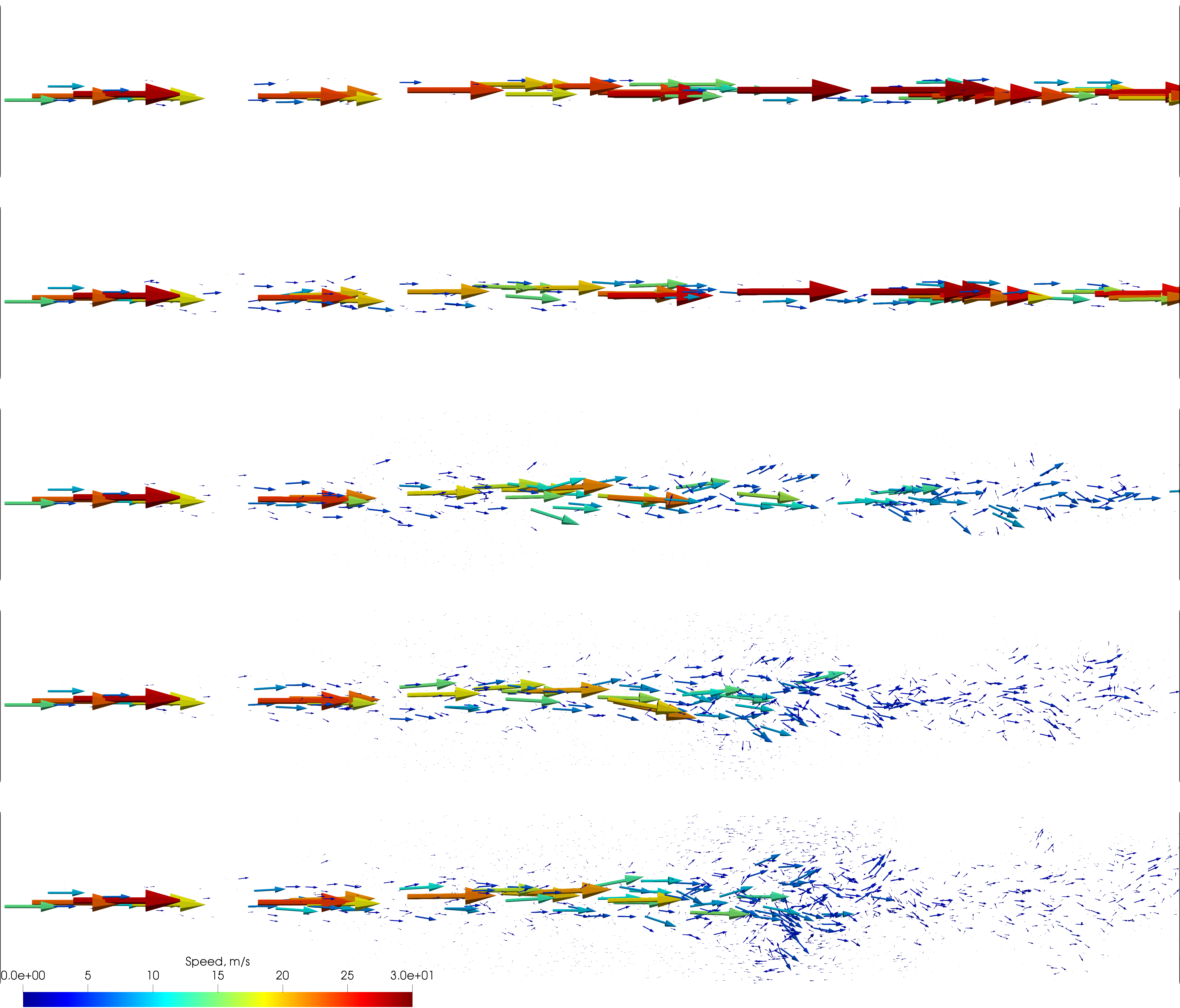}
\caption{Velocity of the flow at times $t=0.01$, $0.02$, $0.03$,
  $0.04$ and $0.05$ s (top to bottom).}
\label{fig:velocity_t0.01-0.05}
\end{figure}

As Reynolds first observed in his experiment \citep{Rey83}, the
development of turbulence is characterized by spontaneous
disintegration of laminar flow into chaotic motions across multiple
scales. Thus, in Figure~\ref{fig:velocity_t0.01-0.05} we present the
lateral view of the pipe, with velocity displayed as a vector field at
times $t=$ 0.01, 0.02, 0.03, 0.04 and 0.05. The direction of arrows in
Figure~\ref{fig:velocity_t0.01-0.05} correspond to the direction of
flow velocity , while their length and color-coding match the flow
speed in that location. As we can see, the initially laminar jet
spontaneously transitions into chaotic flow on the time scale of 0.05
seconds. This is similar to what we observed in our recent works
\citep{Abr22,Abr23,Abr24,Abr25}.

\begin{figure}
\includegraphics[width=\textwidth]{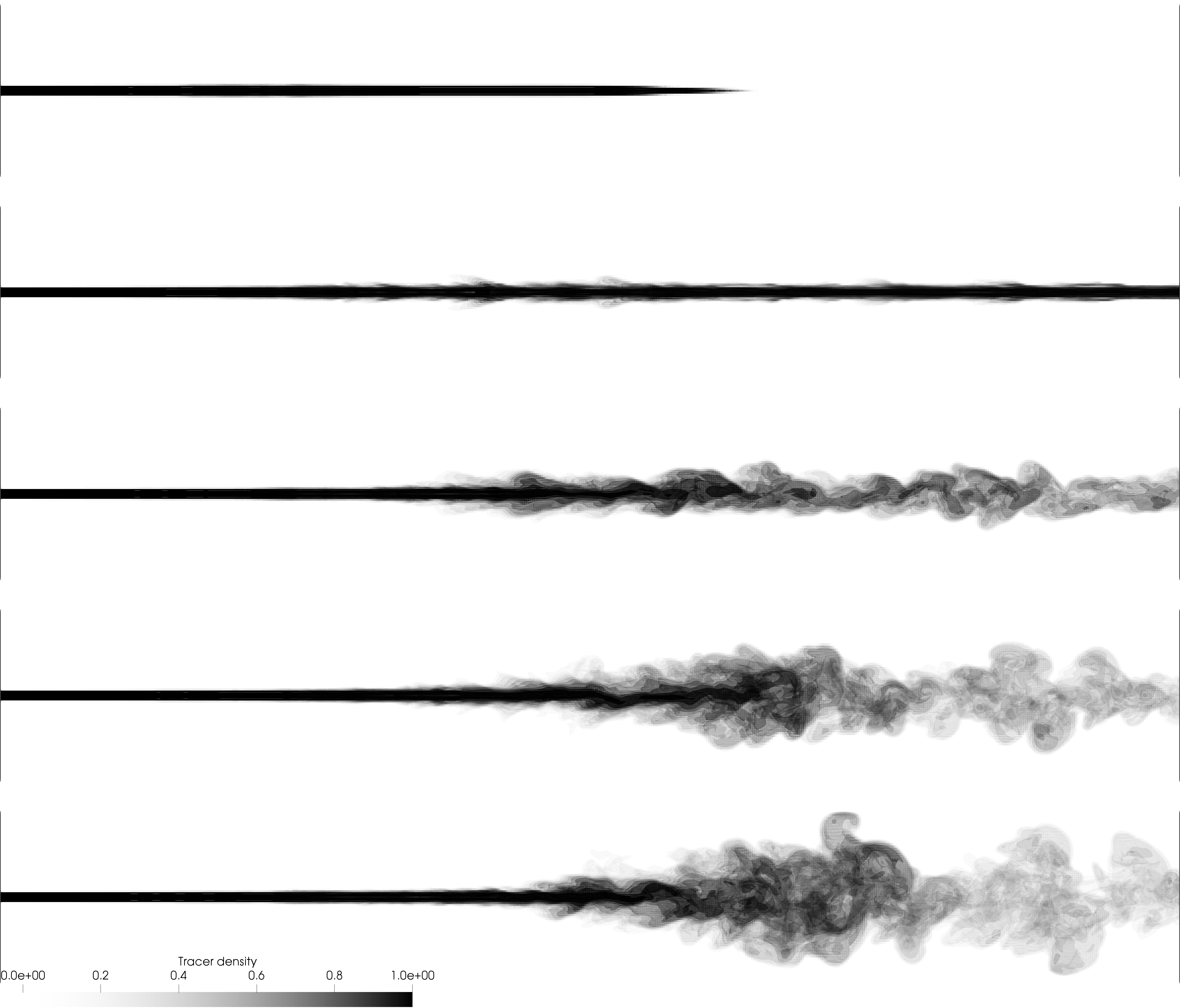}
\caption{The density of the passive tracer at times $t=0.01$, $0.02$,
  $0.03$, $0.04$ and $0.05$ s (top to bottom).}
\label{fig:tracer_t0.01-0.05}
\end{figure}

While the velocity plots are quite illustrative, in practical
observations the onset of turbulence is usually detected via a passive
tracer -- that is, a visually detectable substance, which is
deliberately added to a gas or a liquid. This substance is chosen to
be physically and chemically inert, to avoid interference with the
dynamics of the flow. Here, we simulate the propagation of the passive
tracer using the corresponding equation
\begin{equation}
\parderiv{(\rho s)}t+\nabla\cdot(\rho s\BV u)=0,
\end{equation}
where $s=s(t,\BV x)$ is the density of the tracer substance. We set
the inlet boundary condition for $s$ to be of the Dirichlet type, with
$s=1$ within 0.2 cm distance from the longitudinal axis, and zero
otherwise (such that the initial diameter of the resulting streak is
0.4 cm). In Figure~\ref{fig:tracer_t0.01-0.05} we again present the
lateral view of the pipe, where the density of the passive tracer is
displayed at times $t=$ 0.01, 0.02, 0.03, 0.04 and 0.05. With help of
\textit{ParaView}, the plots in Figure~\ref{fig:tracer_t0.01-0.05} are
produced via level surfaces of the tracer density, distributed
logarithmically from 0.01 to 1, with the opacity of each level surface
set proportionally to the corresponding value of $s$. The logarithmic
distribution of level surfaces is chosen to mimic the logarithmic
sensitivity of a human eye, for a more natural visualization. Again,
we can see that the initially laminar streak of the passive tracer
spontaneously dissolves into the surrounding flow, as typically
observed in experiments (e.g. \citet{Rey83}, p.~942, Fig.~4).

\subsection{Properties of fully developed turbulent flow}

\begin{figure}
\includegraphics[width=\textwidth]{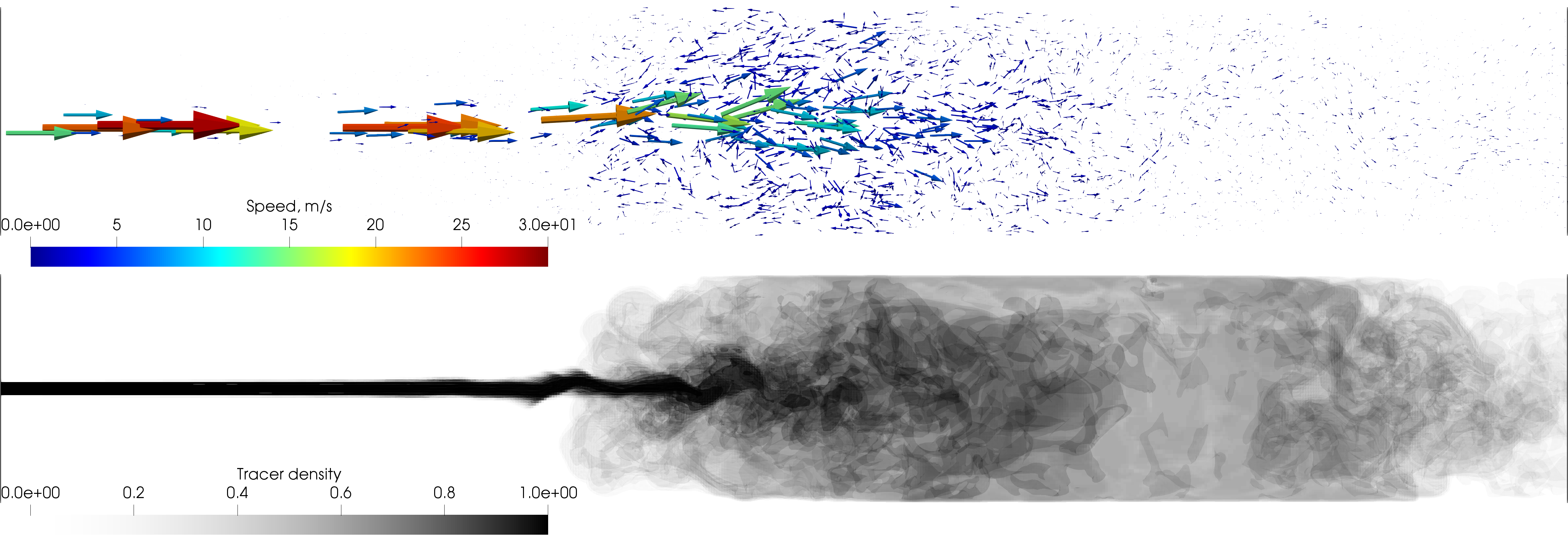}
\caption{Velocity of the flow (top) and density of the passive tracer
  (bottom) at time $t=0.15$ s.}
\label{fig:velocity_tracer_t0.15}
\end{figure}

In the current computational setting, we observed that turbulent flow
fully develops at the elapsed time $t=$ 0.15 s (that is, its
statistical properties do not appear to change noticeably
thereafter). Thus, in Figure~\ref{fig:velocity_tracer_t0.15} we show
velocity of the flow and density of the passive tracer at time $t=$
0.15 s, in the same fashion as in
Figures~\ref{fig:velocity_t0.01-0.05} and \ref{fig:tracer_t0.01-0.05}
above. Observe that velocity has largely same pattern as at $t=$ 0.05
s, except that the complete breakdown of the jet occurs somewhat
closer to the inlet (approximately in the middle of the pipe, or
$\sim$ 24 cm from the inlet). The density of the passive tracer
exhibits matching behavior; namely, it has laminar structure roughly
for the first third of the length of the pipe, and then rather
abruptly breaks down into a partially transparent cloud, inside which
more dense structures are visible. Such behavior appears to be similar
to what was observed by Reynolds,\footnote{\citet{Rey83}, page 942:
  \textit{``\dots the colour band would all at once mix up with
    surrounding water, and fill the rest of the tube with a mass of
    coloured water\dots{} On viewing the tube by the light of an
    electric spark, the mass of colour resolved itself into a mass of
    more or less distinct curls, showing eddies''}} even though he
experimented with turbulent flow of water, rather than air.

\begin{figure}
\includegraphics[width=\textwidth]{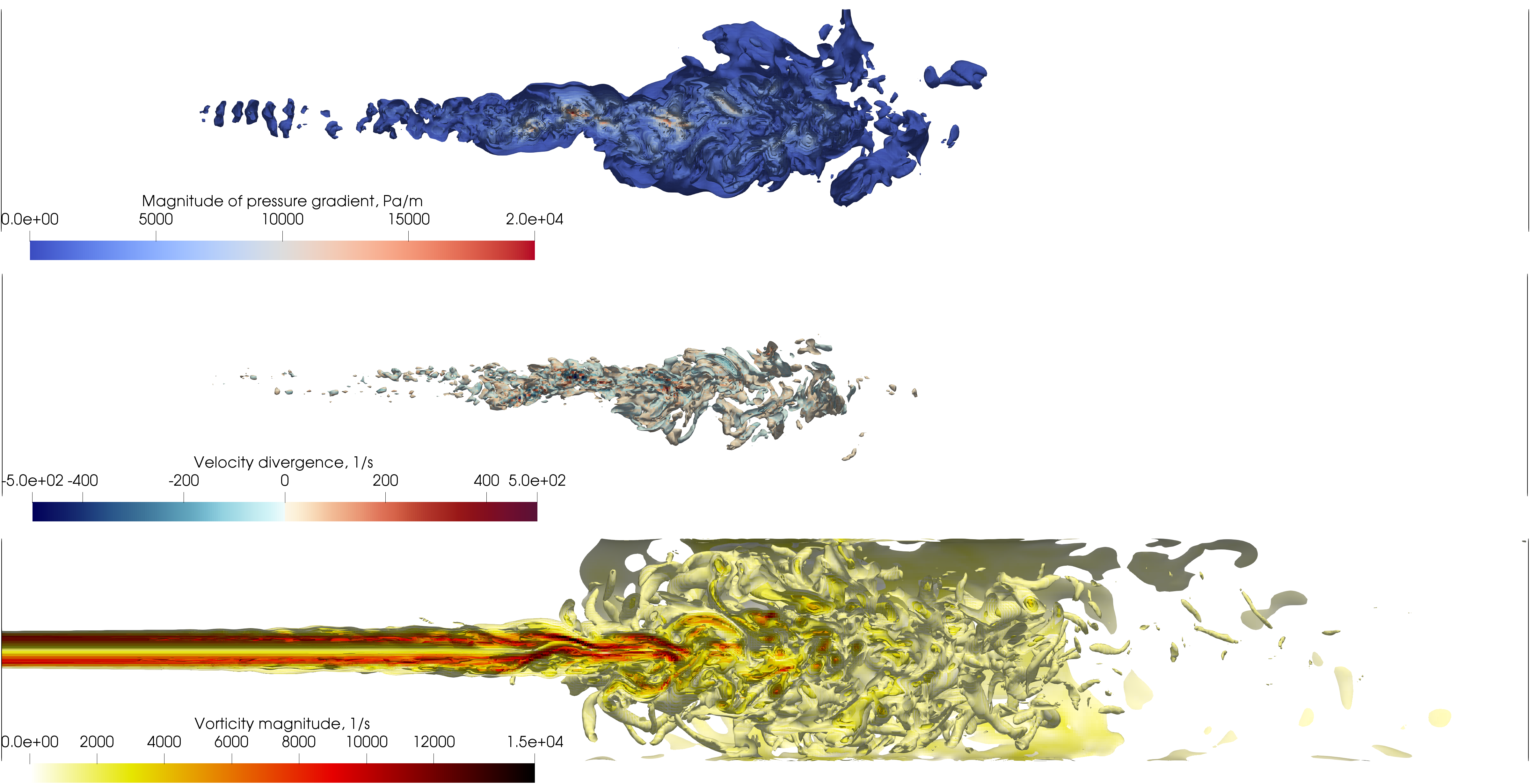}
\caption{Magnitude of the pressure gradient (top), divergence of the
  velocity (middle) and magnitude of the vorticity (bottom) at
  time $t=0.15$~s.}
\label{fig:gradP_divU_vorticity_t0.15}
\end{figure}

In Figure~\ref{fig:gradP_divU_vorticity_t0.15} we show the magnitude
of pressure gradient $\|\nabla p\|$, the divergence of flow velocity
$\nabla\cdot\BV u$, and compare the latter with the magnitude of
vorticity $\|\nabla\times\BV u\|$. All three plots are made as
follows: first, a set of uniformly (rather than logarithmically, as
was the case for the passive tracer) distributed color-coded level
surfaces is created in \textit{ParaView}. Then, all parts of these
level surfaces which are in front of the lateral plane of the pipe are
deleted, thus revealing the ``intestines'' of the plot.

By design, the weakly compressible system in
\eqref{eq:weakly_compressible} aims to keep the non-dimensional
pressure gradient and the velocity divergence at about $O(\Mach^2)$,
while lacking direct mechanisms to restrict vorticity. Observe that
the magnitude of the pressure gradient reaches, at the extreme, 20
kPa/m, which in the non-dimensional units (that is, after multiplying
by $L/p_0$, with $L=$ 1 cm and $p_0=$ 101.3 kPa) constitutes $\sim$
0.002, that is, clearly below $\Mach^2$.  Thus, the primary goal of
\eqref{eq:weakly_compressible} is achieved -- the pressure gradient in
the momentum equation \eqref{eq:u} does not become large at given Mach
number. Additionally, observe that such extreme values of the pressure
gradient are reached only in a few localized spots in the domain,
whereas its bulk values are clearly much lower. In particular, the
average value of $\|\nabla p\|$ over the domain is
\begin{equation}
\frac 1V\int_D\|\nabla p\|\dif\BV x=210\text{ Pa}/\text{m},
\end{equation}
where $V$ signifies the volume of the pipe, and which in the
non-dimensional units totals $2.1\cdot 10^{-5}$, that is, much lower
than $\Mach^2$.

For the divergence of velocity, observe that while its total scale is
between $\pm 500$ s$^{-1}$, such extremal values manifest only in a
few localized spots (just as for the pressure gradient above), whereas
most values remain within $\pm 50$ s$^{-1}$, according to the colors
of the level surfaces. The average value of $|\nabla\cdot\BV u|$ over
the domain is
\begin{equation}
\frac 1V\int_D|\nabla\cdot\BV u|\dif\BV x=0.932\text{ s}^{-1}.
\end{equation}
In non-dimensional units, we have to multiply the result by $L/U\approx
3\cdot 10^{-4}$ s, which yields $\tilde\nabla\cdot\BV{\tilde u}\sim\pm
1.5\cdot 10^{-2}$, while the non-dimensional domain average of its
absolute value is $\sim 3.2\cdot 10^{-4}$. These magnitudes are
generally in agreement with $O(\Mach^2)\sim 10^{-2}$, as intended.

By contrast, the norm of vorticity is about two orders of magnitude
larger than that of divergence throughout most of the domain; in
particular, its domain average is
\begin{equation}
\frac 1V\int_D\|\nabla\times\BV u\|\dif\BV x=408\text{ s}^{-1}.
\end{equation}
This indicates that the overall structure of the flow is weakly
compressible, and largely rotational.

In our recent works \cite{Abr22,Abr23,Abr24} we found that, in an
inertial flow, the density and temperature variables could develop
small scale fluctuations of as much as $\pm 60$\% of their background
values, which, of course, is unrealistic for low Mach number flows at
normal conditions. We speculated that this happens because, in a
realistic flow, the pressure gradient responds to large density and
temperature fluctuations (thus allowing the flow to ``compensate'' for
the latter), whereas in inertial flow the pressure gradient is zero.

\begin{figure}
\includegraphics[width=\textwidth]{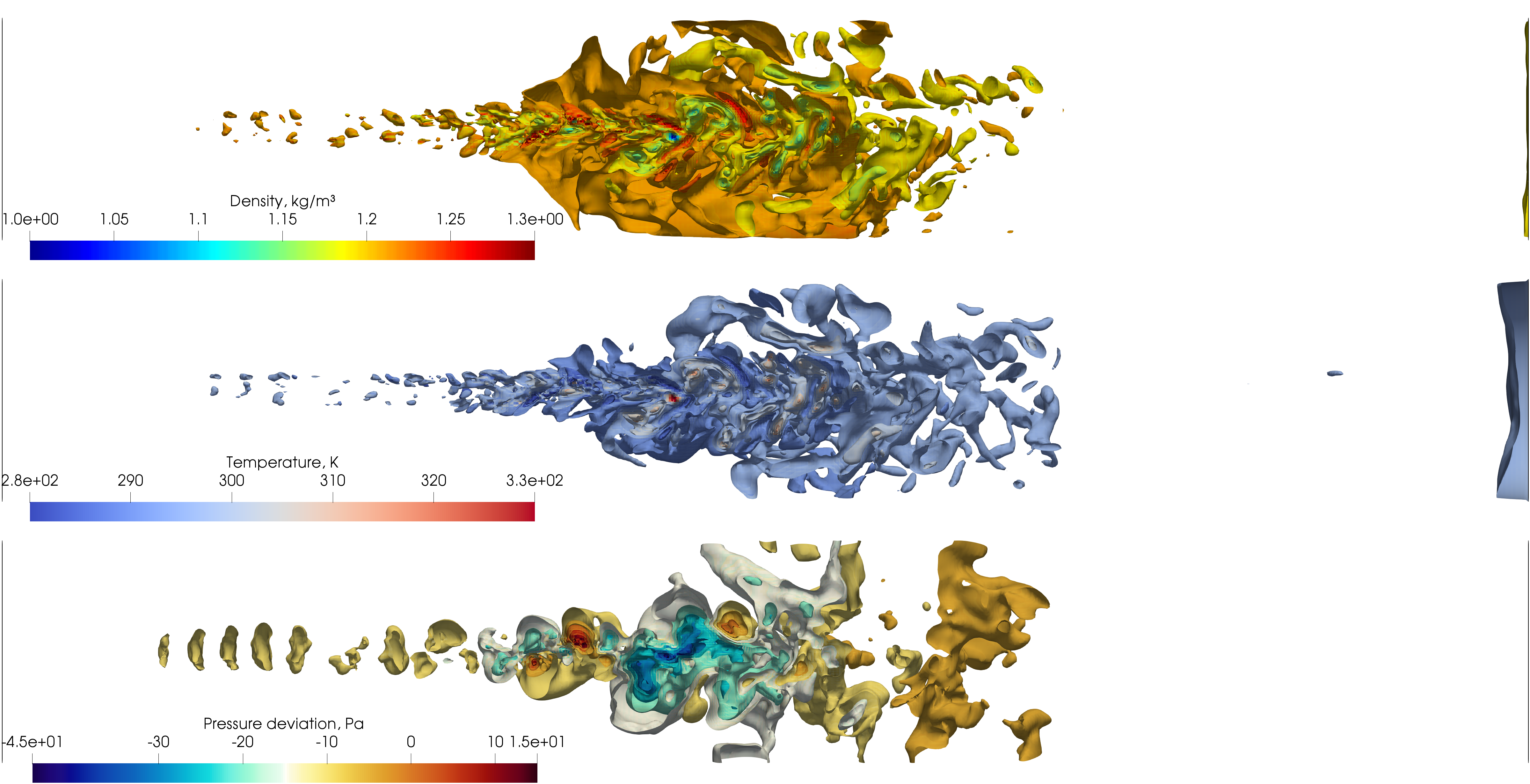}
\caption{Density (top), temperature (middle) and pressure deviation
  (bottom) at time $t=0.15$ s.}
\label{fig:rho_T_p_t0.15}
\end{figure}

Since in the weakly compressible model \eqref{eq:weakly_compressible}
the pressure is no longer constant, here we examine whether this
modification has a beneficial effect on the density and temperature
fluctuations. In Figure~\ref{fig:rho_T_p_t0.15} we show the snapshots
of density, temperature and the deviation of pressure from its
background state of 101.3 kPa, produced in the same fashion as those
of divergence and vorticity above. First, we note that the pressure
deviations lie, at the extreme, between $-45$ Pa and $+15$ Pa; given
the background state of $101.3$ kPa, this constitutes $-0.045$\% and
$+0.015$\%, respectively, which indicates that the flow is still
``almost'' inertial. Yet, we can observe vast improvement in the
fluctuations of density and temperature of the flow. For example,
temperature now varies between $280$ K and $330$ K at the extreme
(which for the background state of $293.15$ K constitutes only
$-4.5$\% and $+13$\%, respectively) and even then, these extreme
deviations are observed only at a few localized spots, while the
majority of deviations are within the range of several degrees K.
Moreover, the domain average of the temperature deviation from its
background state of 293.15 K is
\begin{equation}
\frac 1V\int_D|T-293.15|\dif\BV x=0.716\text{ K},
\end{equation}
which constitutes 0.24\% of the background state itself. A similar
trend is observed for the density fluctuations. We conclude that the
weakly compressible model in \eqref{eq:weakly_compressible} produces
notably more realistic behavior of density and temperature, than the
inertial flow model in our recent works, while retaining the essential
turbulent dynamics of the flow.

In addition to the snapshots of various properties of the flow,
displayed above, we compute time averages of Fourier spectra of the
streamwise kinetic energy. Each time average is computed in the same
fashion as in our recent works \citep{Abr22,Abr23,Abr24,Abr25}, within
a cuboid region of 24 cm in length, and of 1$\times$1 cm$^2$ in
cross-section. Six of these regions were placed inside the pipe, with
their axes coinciding with the axis of the pipe, at 0, 4, 8, 16, 20
and 24 cm offsets from the inlet (such that the first region started
at the inlet, while the sixth one ended at the outlet). In each
region, the Fourier spectrum of the kinetic energy was computed as
follows: first, the kinetic energy of the $x$-component of velocity,
$E_x=u_x^2/2$, was averaged over the cross-section of the region, thus
becoming the function of the $x$-coordinate only. Then, the linear
trend was subtracted from the result in the same manner as was done by
\citet{NasGag} and also in our recent works
\citep{Abr22,Abr23,Abr24,Abr25}, to ensure that there was no sharp
discontinuity between the energy values at the boundaries of the
region. Finally, the one-dimensional discrete Fourier transformation
was applied to the result. The subsequent time-averaging of the
modulus of the Fourier transform was computed in the time interval
$0.15\leq t\leq 0.25$ seconds of the elapsed model time.

\begin{figure}
\includegraphics[width=\textwidth]{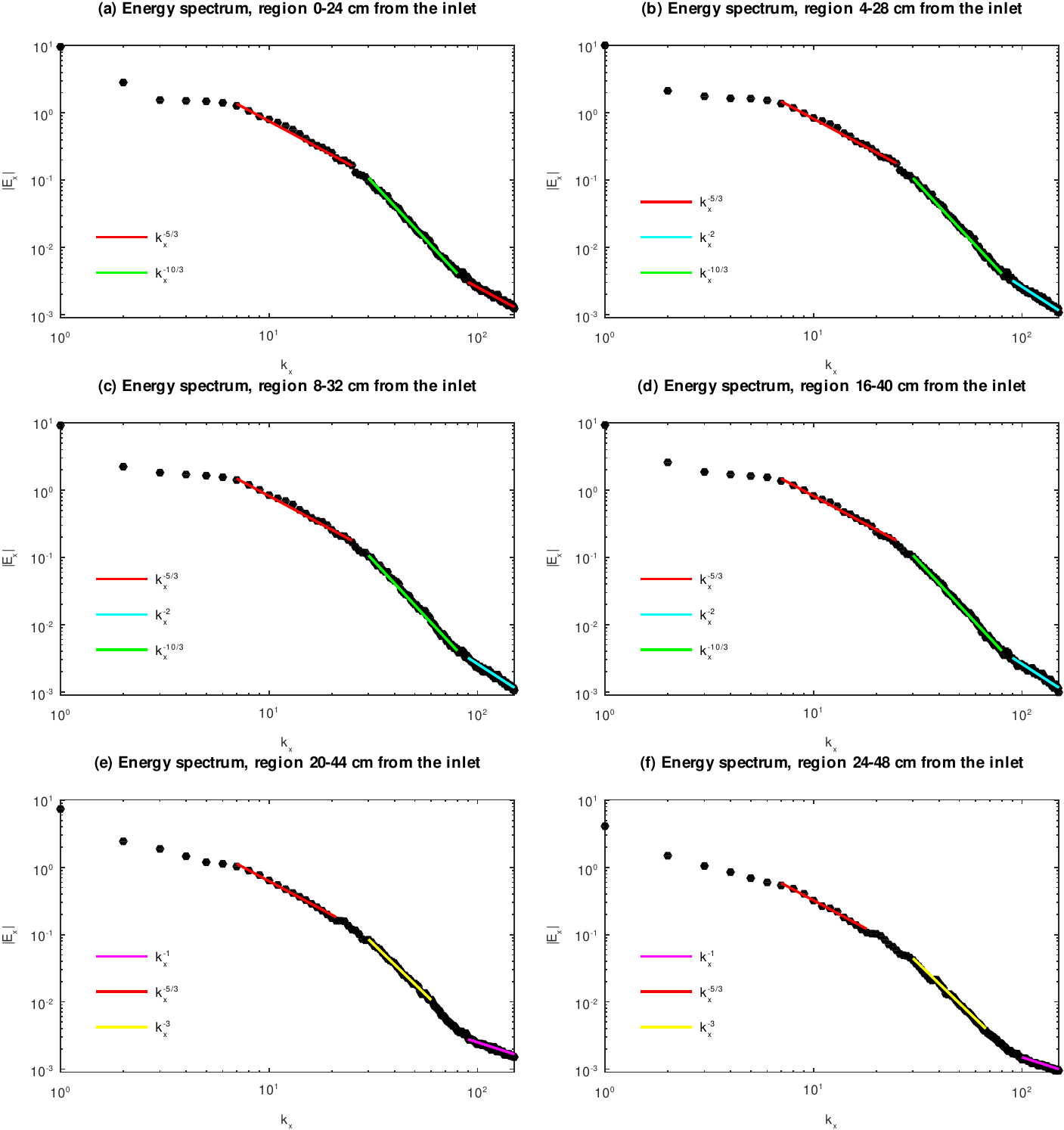}
\caption{The Fourier spectra of the kinetic energy, computed within
  different regions, situated at ({\bf a}) 0--24, ({\bf b}) 4--28,
  ({\bf c}) 8--32, ({\bf d}) 16--40, ({\bf e}) 20--44 and ({\bf f})
  24--48 cm, counting from the inlet. The slope lines $k_x^{-1}$,
  $k_x^{-5/3}$, $k_x^{-2}$, $k_x^{-3}$ and $k_x^{-10/3}$ are given for
  reference.}
\label{fig:energy_spectra}
\end{figure}

We show the time averages of the kinetic energy spectra in
Figure~\ref{fig:energy_spectra}. Observe that, like in our recent
works \citep{Abr22,Abr23,Abr24,Abr25}, these spectra generally retain
a power scaling structure. However, unlike our previous results (in
particular, \citep{Abr24}), this structure appears to be somewhat
different from what we observed before. While the Kolmogorov scaling
$k_x^{-5/3}$ is still present at moderate scales (for $k_x$ roughly
between 8 and 25), observe that at small scales ($k_x$ between 30 and
80) the slope of the plot corresponds to $k_x^{-10/3}$ for the first
four regions, and to $k_x^{-3}$ for the last two regions.
Additionally, at very small scales ($k_x>100$) the power scaling
varies between $k_x^{-2}$ and $k_x^{-1}$, which has not been observed
in our previous works. Additionally, Kolmogorov's $k_x^{-5/3}$ scaling
at moderate scales is present in all measurement regions, whereas in
our recent work \citep{Abr24} it disappeared in the regions which
approached the outlet. Given that the only difference between the
weakly compressible system in \eqref{eq:weakly_compressible} and the
fully inertial flow in \citep{Abr22,Abr23,Abr24,Abr25} is a different
treatment of the pressure variable, we have to conclude that the power
structure of the kinetic energy spectrum is rather sensitive to the
behavior of the pressure variable (recall from
Figure~\ref{fig:rho_T_p_t0.15} that pressure fluctuations constitute
only a tiny fraction of its background value).

\section{Summary}
\label{sec:summary}

In the current work, we propose a new approach for treating the
pressure variable, which we refer to as weakly compressible flow.
Instead of setting the pressure so as to preserve the divergence of
velocity along the streamline (as is done for incompressible flow), we
instead formulate the pressure equation so that the divergence is
linearly damped towards zero at an appropriate time rate. Our new
approach is a compromise between compressible, incompressible and
inertial flows, in the sense that it controls the pressure gradient at
low Mach number, while simultaneously allowing the flow to be
compressible, and has a more realistic density and temperature
behavior than inertial flow. We find that, in the presence of the mean
field effect of the intermolecular potential, our new model of weakly
compressible flow spontaneously develops turbulent motions and power
structures in its kinetic energy spectra, just like the inertial flow
model did in our recent works \citep{Abr22,Abr23,Abr24,Abr25}.

The results of the current work are rather encouraging. Clearly, our
main finding here is that the spontaneous development of turbulence
via the mean field effect of the intermolecular potential is robust
under different treatments of the pressure variable; indeed,
turbulence reliably develops in the same realistic fashion
irrespectively of whether the flow is inertial (which usually happens
at large scales, such as planetary scales) or weakly compressible
(which is more realistic at small scales). This is in agreement with
experiments and observations, where turbulent motions manifest
similarly across a broad range of spatial scales, from centimeters to
thousands of kilometers. This is also in sharp contrast with the
conventional equations of fluid mechanics, both compressible and
incompressible, where the spontaneous development of turbulence has
not ever been observed despite overwhelming research efforts spanning
more than a century.

The next step in this direction would be the development of an
appropriate pressure model from fundamental physical principles,
rather than technical considerations. Observe that, in the current
work, the pressure treatment is ``engineered'', rather than derived --
the main argument is that we need to somehow keep the pressure
gradient under control at low Mach number, and the proposed model is
artificially tailored for that purpose. While at the current stage it
is nevertheless a good result (at least in the particular case studied
here), a systematic treatment of pressure, grounded in physical
foundation, is essential for our turbulence model to provide reliable
predictions in a broad range of real-world scenarios.

\ack This work was supported by the Simons Foundation grant \#636144.

\end{document}